\documentclass[10pt,conference]{IEEEtran}
\IEEEoverridecommandlockouts
\usepackage{cite}
\usepackage{amsmath,amssymb,amsfonts}
\usepackage{algorithmic}
\usepackage{graphicx}
\usepackage{comment}
\usepackage{caption}
\usepackage{cite}
\usepackage{subcaption}
\usepackage{textcomp}
\usepackage{xcolor}
\def\BibTeX{{\rm B\kern-.05em{\sc i\kern-.025em b}\kern-.08em
    T\kern-.1667em\lower.7ex\hbox{E}\kern-.125emX}}
\begin{document}

\title{Noise Mitigation with Delay Pulses in the IBM Quantum Experience\\
\thanks{This work was supported by the Natural Sciences and Engineering Research Council of Canada through its Undergraduate Student Research Award and Discovery programs (RGPIN-2020-04328).}
}

\author{\IEEEauthorblockN{Sam Tomkins}
\IEEEauthorblockA{\textit{Department of Physics and Astronomy} \\
\textit{University of Victoria}\\
Victoria, Canada \\
stomkins@uvic.ca}
\and
\IEEEauthorblockN{Rog\'{e}rio de Sousa}
\IEEEauthorblockA{\textit{Department of Physics and Astronomy} \\
\textit{University of Victoria}\\
Victoria, Canada \\
rdesousa@uvic.ca}
}

\maketitle

\begin{abstract}
One of the greatest challenges for current quantum computing hardware is how to obtain reliable results from noisy devices. A recent paper [A. Kandala {\it et al.}, Nature {\bf 567}, 491 (2019)] described a method for injecting noise by stretching gate times, enabling the calculation of quantum expectation values as a function of the amount of noise in the IBM Q devices. Extrapolating to zero noise led to excellent agreement with exact results. Here an alternative scheme is described that employs the intentional addition of identity pulses, pausing the device periodically in order to gradually subject the quantum computation to increased levels of noise. The scheme is implemented in a one qubit circuit on an IBM-Q device. It is determined that this is an effective method for controlled addition of noise, and further, that using noisy results to perform extrapolation can lead to improvements in the final output, provided careful attention is paid to how the extrapolation is carried out.
\end{abstract}


\section{Introduction}
The progress in experimental realization of quantum computers has been remarkable. Several compa­nies such as D-Wave, IBM, and Rigetti are now providing cloud­-based access to small­ scale quantum processors based on superconducting hardware, with many others such as Google and AliBaba planning to launch their quantum cloud platform soon \cite{castel}. All these devices are classified in the Noisy Intermediate­ Scale Quantum (NISQ) category, because they are noisy and do not have enough qubits to perform quantum er­ror correction \cite{preskill}. Noise greatly reduces the capacity of quantum computers to solve problems, and washes out their “quantum advantage”. 

The IBM Q devices are based on the gate model of quantum computing. An open-source quantum development kit called Qiskit \cite{qiskit} allows users to construct and illustrate circuits, execute them on IBM quantum computers through the cloud, and visualize and interpret the results. Qiskit also provides a quantum simulator, which calculates the exact outcome expected from a circuit, for comparison to that of a real quantum device.

The ability to test noisy quantum computers has led to the emergence of an exciting new area of research: The development of protocols for noise mitigation [4,5]. In Richardson extrapolation \cite{richardson} one measures expectation values at different levels of noise and extrapolates to the zero noise limit. This was demonstrated in IBM-Q's five qubit \emph{Yorktown} device by running the same quantum algorithm (finding the ground state energy of a molecule) as a function of longer quantum gate times. Longer gate times implied the qubits were subjected to more decoherence, enabling a plot of quantum expectation values as a function of the amount of noise. Extrapolating to zero noise led to results that agreed with exact solutions \cite{kandala}. While theory shows that Richardson extrapolation works for $T_1$ dominated decoherence \cite{temme}, it is not known whether it works when additional low frequency $T_2$ processes are present. Moreover, it is not known how well it works for different quantum algorithms. A key requirement is the availability of a ``knob'' to tune the amount of noise. Below the use of identity delay pulses is proposed as an alternative means to perform noise mitigation. 

In \cite{kandala} noise was added to the system by stretching gate times, meaning that the pulse sent through the qubit to perform each gate operation is lengthened. This showed two important positive results. Firstly, the stretching of quantum gates can effectively add error to the system in a controlled manner. Secondly, extrapolation of increasingly noisy results is successful in improving the result of the computation. These two results provide grounds on which our method of adding delay gates can be evaluated: the viability of using identity gates as a controlled knob to add noise, and the ability of extrapolation to reduce noise when this technique is used.

\section{One Qubit Evolution Subject to Intrinsic and Extrinsic Noise}

\subsection{Target Algorithm}
The target algorithm that will be improved by zero-noise extrapolation is a series of gates that take a qubit from the pure $|0\rangle$ state to the pure $|1\rangle$ state in 30 distinct steps. This is implemented by the recursive equation \ref{eqn:algorithm}, where the 30 steps correspond to $j = 0, 1, 2, . . . 30$.  Each $R$ indicates a rotation by the sub-scripted angle around the super-scripted axes, and $U_0$ is the identity operation, ending the recursivity.
\begin{equation}
\label{eqn:algorithm}
U_{j+1} = R^Z_{4(j+1)\pi/30}R^X_{(j+1)\pi/30}R^X_{-j\pi/30}R^Z_{-4j\pi/30}U_{j}.
\end{equation}

In Qiskit, the sequence of programmed gates is processed by a transpiler whose goal is to optimize the circuit and express gates such as $R^{Z}_{\alpha},R^{X}_{\beta}$ in terms of native gates. However, for this work the transpiler has to be turned off, because it automatically removes delay gates. This is not a problem for implementing Eq. \ref{eqn:algorithm}, as $R^{Z}_{\alpha}$ is implemented in Qiskit by the native operation $u_1(\alpha)$, and the $R^{X}_{\beta}$ gate is implemented by $u_3(\beta, -\pi/2,\pi/2)$ \cite{ibmq}.

\begin{figure}[htbp]
\centerline{\includegraphics[width=0.82\linewidth]{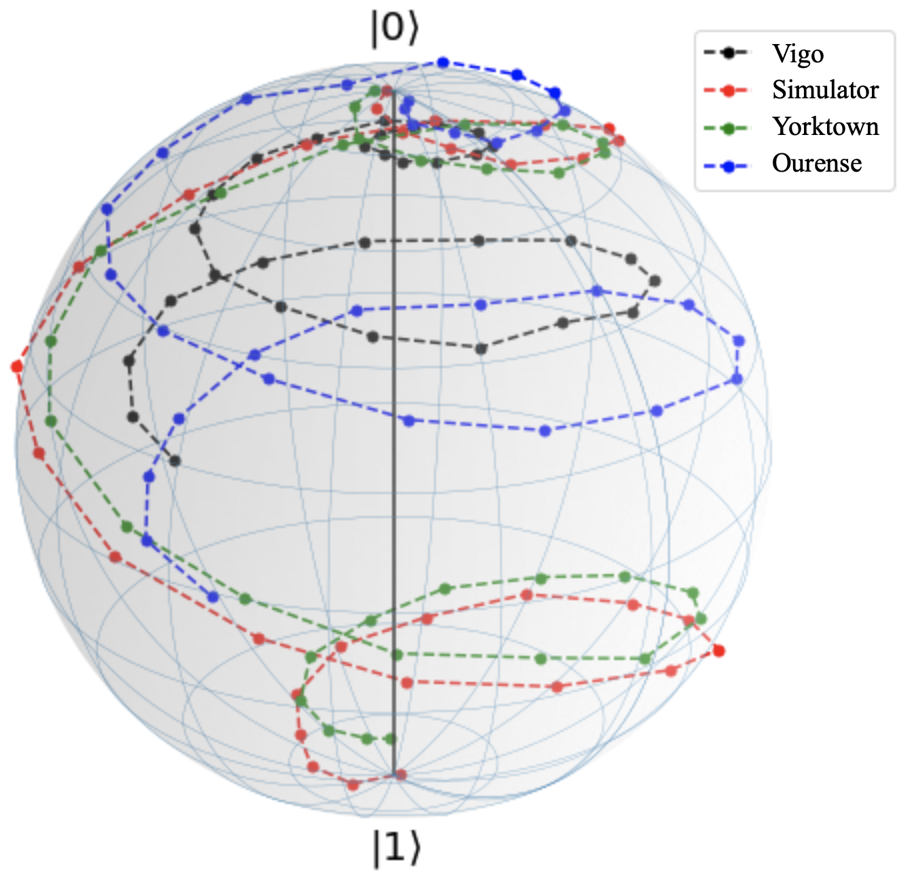}}
\caption{Trajectories produced by equation \ref{eqn:algorithm} computed using IBM Q devices Yorktown, Vigo, and Ourense, as well as the exact result determined by simulation.}
\label{fig:baseline}
\end{figure}

The one-qubit algorithm described by Eq. \ref{eqn:algorithm} was executed on all three of the IBM-Q 5 qubit devices available at the time (Yorktown, Ourense, and Vigo). The results of these computations produce the trajectories of Figure \ref{fig:baseline}, with the exact result determined using the simulator included for comparison. These results motivated the use of the Yorktown device moving forward. The trajectories from Ourense and Vigo stray so far from the expected curve that any correction technique is sure to be useless, whereas the Yorktown trajectory has the potential to be corrected closer to the exact result.

\subsection{Method for Noise Injection}
\begin{figure}[htbp]
\centerline{\includegraphics[width=\linewidth]{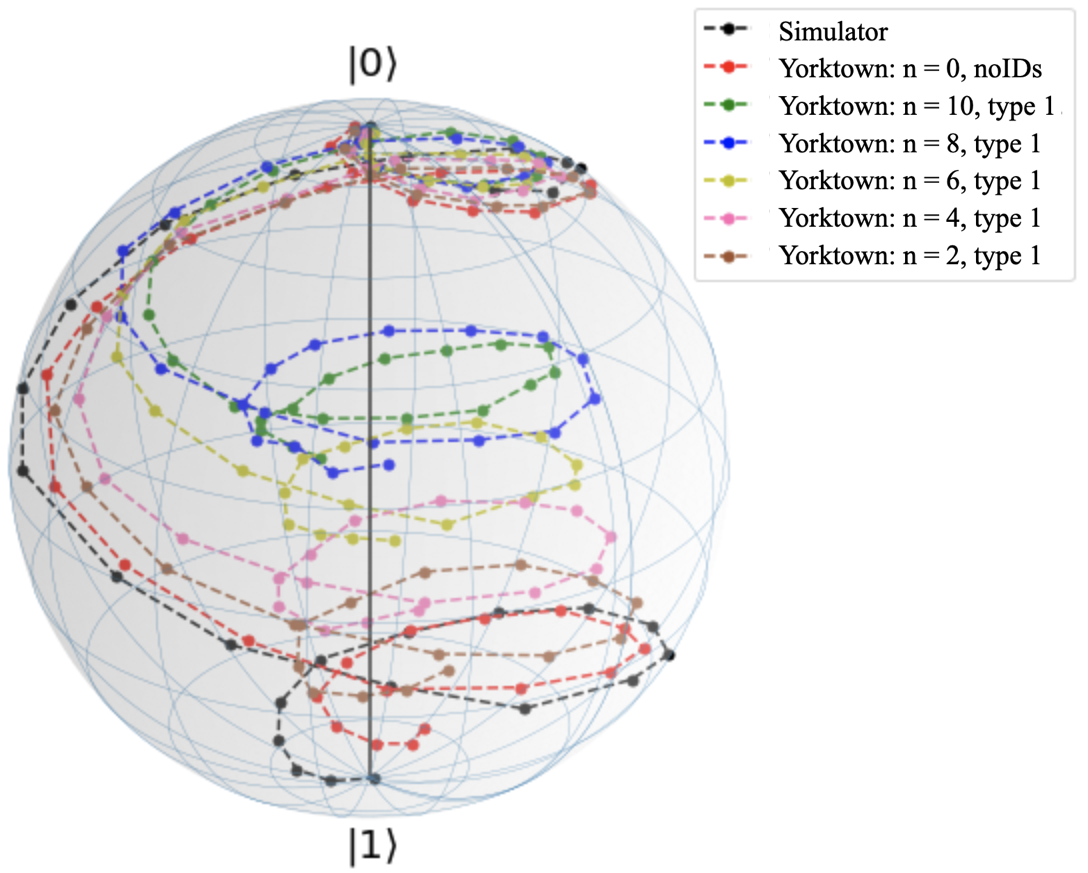}}
\caption{Trajectories produced by intentionally adding noise using the type 1 method (eqn. \ref{eqn:type1}), computed using IBM Q Yorktown device.}
\label{fig:type1}
\end{figure}

In order to perform extrapolation and reduce noise it is necessary to first intentionally add noise to the system. As outlined previously, this is achieved by placing identity gates in the circuit. Each additional identity gate extends the execution time by pausing the computation for 70ns. Utilizing many of these gates can make the algorithm take significantly longer to run, in turn increasing the amount of noise. The choice of where to add these gates in the algorithm is an important consideration. For this work 3 potential ways to include identity gates were considered. For the purpose of explanation they will be referred to as type 1, 2, and 3. Type 1 consists of placing a fixed number, $n$, of identity gates between every gate in the circuit (as below). Type 2 is a method where all $n$ identity gates are added only at the end of the circuit before measurement (as below). Finally, type 3 is a combination of types 1 and 2, where n identity gates are placed every 5 regular gates, corresponding to one set between every step of the algorithm (as below). Since $n$ refers to the number of identity gates in a single set, it is necessary to choose a different value for each method so that the total number is comparable across the entire circuit. This ensures that a similar amount of noise is introduced in each case.
\begin{subequations}
	\begin{align}
	\label{eqn:type1}
	R^Z_{4(30+1)\pi/30}(Id)^n& \ldots R^Z_{-4(1)\pi/30}(Id)^n,\\[1em]
	\label{eqn:type2}
	U_{30}U_{29}& \ldots U_{1}U_0(Id)^n,\\[1em]
	\label{eqn:type3}
	U_{30}(Id)^nU_{29}(Id)^n& \ldots U_1(Id)^nU_0(Id)^n.
	\end{align}
\end{subequations}
The purpose of adding identity gates is to increase noise in a controlled manner. These three methods were considered to determine which best met this criteria. The trajectories produced using type 1 noise injection are shown in Figure \ref{fig:type1}. The other two methods were considered but are not shown; type 2 leads to inconsistencies in the deviation from the exact curve as more gates are added. Meanwhile, type 3 results in non-smooth behaviour in some of the curves. This leaves type 1 as the preferred method for introducing error, and all further work is based on noise introduced in this way.

Before discussing extrapolation there are some important conclusions that can be drawn from these results. The gradually increasing noisy trajectories of Figure \ref{fig:type1} show that the use of identity gates acts as a knob to smoothly add noise in a quantum computation. The type 1 method employed increases this noise in a consistent and controlled manner. This is seen is the continual deviation from the expected trajectory as more identity gates are added. These results provide a solid foundation on which to begin extrapolating and attempting to correct the control result to the zero-noise regime.

\section{Extrapolation to Zero Noise}
In the strictly mathematical definition extrapolation refers to approximating values outside the range of where they are known based on the relationship between the known values and a controllable variable \cite{brezinski}. In the case presented here the measured values are the noisy trajectories obtained previously, and the controllable variable is the amount of identity gates added.The objective of extrapolation is to approximate a trajectory with less noise than the case where no identity gates were used. Two methods are explored for extrapolation; a simple linear approach and the more complicated Richardson process \cite{richardson}.

Generally, deciding on an extrapolation procedure requires an understanding of the relationship between the data and the external variable. In the context of this work an exact relationship between adding identity gates and increasing noise is not entirely known, so it is impossible to define an extrapolation method optimized for this application. Instead, the noisy results must be carefully considered to determine what makes sense for this particular case.

\subsection{Linear Extrapolation}
A reasonable starting point is to assume that the trajectories deviate linearly with the number of identity gates used. This assumption allows a simple linear extrapolation to be applied. This technique uses a linear regression approach to produce a best fit line through the known data. An unknown value is then approximated as a point on this line outside of the range where data is known. Further details can be found in \cite{brezinski}.

\begin{figure}[htbp]
	\centerline{\includegraphics[width=\linewidth]{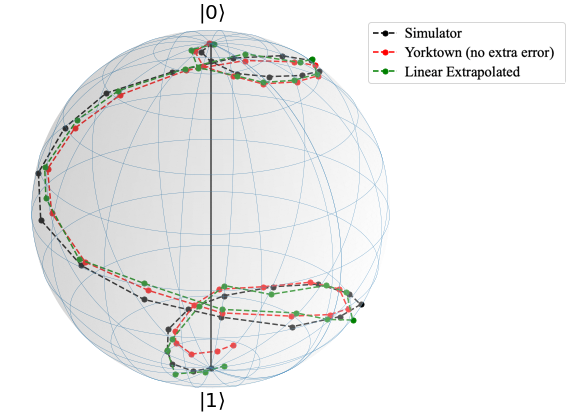}}
	\caption{Trajectory produced using linear extrapolation.}
	\label{fig:linear}
\end{figure}

In order to apply this simple one dimensional technique to the three dimensional trajectories studied each point is decomposed into its x, y, and z coordinates, with extrapolation performed individually along each cardinal axis. This results in 90 iterations of the extrapolation procedure described above; 30 points, each with 3 directions. Every one of these steps uses 11 known data points, with the controllable variable being $n = 0, 1, 2, ...10$, to produce one extrapolated point. The 90 individual results are recombined to form a single extrapolated trajectory. The final outcome is shown as the green curve in Figure \ref{fig:linear}, which is compared to the control case as well as the result of the simulator.

The result of linear extrapolation is promising given the simplicity of the procedure. Throughout the trajectory, and specifically near the final $|1\rangle$ state, there is a significant improvement of the extrapolated (green) curve to match the expected (black) curve. This appears to indicate that the procedure is having the desired effect. However, extrapolation results should always be taken with skepticism, especially those obtained using linear methods. Numerous approximation methods are based on the idea that any function is linear in a small enough window. The problem with this approach in an extrapolation procedure is that this window gets arbitrarily extended without ensuring this approximate linearity continues. When the behaviour of the data is not known continued linearity could be a completely untrue assumption: it is possible that the true trajectory would change drastically as the zero-noise regime is approached, as this is often the case for any function approaching a fundamental limit.

Another consideration when performing linear extrapolation is choosing the value of the controllable variable, in this case the amount of error, corresponding to the zero-noise regime. In the data measured from the real device the control curve, with no extra errror added, corresponds to $n=0$. Of course this curve still has error, so determining where on the extrapolation line the true zero-noise regime begins is somewhat arbitrary. Since the exact final result was known beforehand, a reasonable choice was made to extrapolate until the vertical component of the final point matched its exact value. This corresponded to an extrapolation point where the controllable variable was $n=-0.96$. All other extrapolations were then performed using this value. Doing so produced a good result, but required the exact solution to be known for use in the calibration step to determine the value of $n$ for which zero-noise happens. In more complicated algorithms the exact result may not be known, and this removes the viability of this procedure in such cases.

Knowing the exact result can present additional problems when interpreting an extrapolation. Under these circumstances it is easy to falsely conclude an extrapolation has been successful merely because it becomes closer to what was expected beforehand. In reality the seemingly effective extrapolation may simply be a product of the specific problem being analyzed, or how the procedure was set up. This is especially true in the case here where the extrapolation is, at least partially, based on knowing the exact result. Because of these issues the results of this section are not concluded as a positive improvement, but rather a justification to explore more complex methods.

\subsection{Richardson Extrapolation}
The results of the linear extrapolation are encouraging but the discussion about the shortcomings of such a method leave more to be explored. Such a basic technique is limited in its usefulness, and better results may be obtained through more complicated methods. Building on the work of Kandala et al. \cite{kandala}, the Richardson extrapolation method is also utilized, as this procedure gave excellent results in their work with quantum computing. Richardson extrapolation is a method used to improve the rate at which a sequence of approximations converges to its true value. Equation \ref{eqn:richardson} describes the fundamental procedure, which says that a true value $A^*$ is approximately equal to the difference between two values $A(h/t)$ and $A(h)$, that are measured with different error parameters $h/t$ and $h$ respectively. In this context it is important here that the error parameter scales accordingly between the two measurements. 
\begin{equation}
\label{eqn:richardson}
A^* \approx \frac{t^{k_0}A(h/t)-A(h)}{t^{k_0}-1}
\end{equation}

\begin{figure}[htbp]
	\centerline{\includegraphics[width=\linewidth]{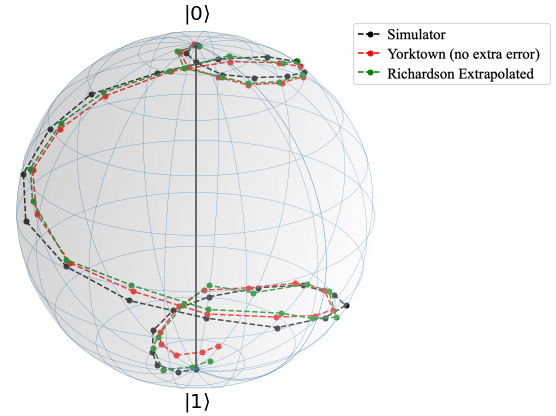}}
	\caption{Trajectory produced using Richardson extrapolation. The numerical procedure for this is an adapted version of Ulerich’s MATLAB script for Richardson extrapolation \cite{ulerich}.}
	\label{fig:richardson}
\end{figure}

As with the linear extrapolation, the x, y, and z coordinates for each of 30 points are the sequences being extrapolated, meaning the method is applied to a total of 90 sequences. The entries of each are the measured values with increasingly more noise added. In the language of equation \ref{eqn:richardson} these are the $A(h)$ values, where the parameter $h$ is related to how many identity gates have been added. After applying the procedure, an improved estimation of the true value is obtained for each sequence. Recombining the results produces the extrapolated trajectory shown in Figure \ref{fig:richardson}.

Superficially this result is very similar to the linear case, with the extrapolated trajectory matching the exact curve much better than the control result, specifically near the end points. While comparable, the result of the Richardson extrapolation has greater significance. This method does not have the linearity issues discussed with the simpler case, meaning the use of this method is more justified with less skepticism required. In order to implement Richardson extrapolation three things must be known; the error parameter ($h$), the scaling factor between step sizes of data ($t$) and the step size behaviours of truncation error (the $k_i$’s). Both $h$ and $t$ are easy to determine: $h$ is essentially the circuit execution time, which scales directly with the number of identity gates added so that $t$ follows directly from this, but the $k_i$ values are not known. Typically these values come from some theoretical model of the problem, but in this case the behaviour is not known well enough and $k_i$’s must be estimated. This can be done using the results measured with different error parameters. This means that there is no prior assumptions being made in order to implement the Richardson method, and the procedure is in no way based on the exact known result. As such the results of this section are much more meaningful, and we can confirm with at least some certainty that a Richardson extrapolation method does improve the results of a quantum computation.

\subsection{Vertical Extrapolation Only}
\begin{figure}[htbp]
	\begin{subfigure}{0.5\textwidth}
		\centering
			\centerline{\includegraphics[width=\linewidth]{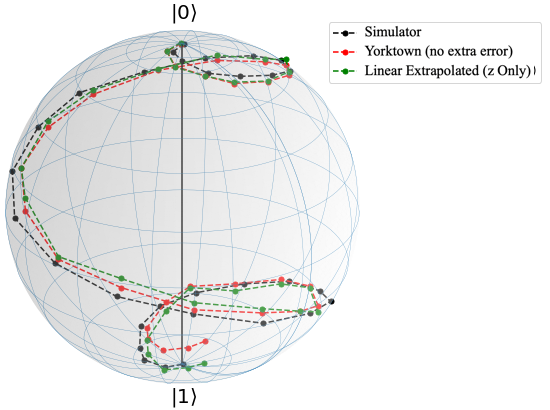}}
		\caption{Linear}
		\label{fig:zonlyLinear}
	\end{subfigure}
	\begin{subfigure}{0.5\textwidth}
	\centering
	\centerline{\includegraphics[width=\linewidth]{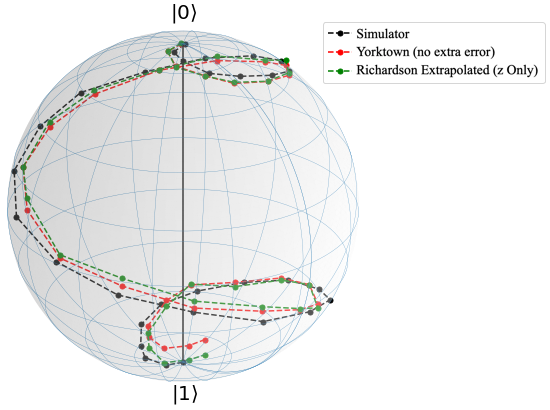}}
	\caption{Richardson}
	\label{fig:zonlyRich}
	\end{subfigure}
	\caption{Trajectories produced using linear and Richardson extrapolation, applied only in the z direction.}
	\label{fig:zonly}
\end{figure}

In the previous two sections an improved trajectory was produced by extrapolating each point along all three cardinal axes. While the results of this method are quite good, there is reason to consider only performing extrapolation in the vertical (z) direction, leaving the other two components of each point unchanged. Since the type of noise being studied results in the decay of an excited qubit to its lower energy state, it is more likely to occur when there is a higher population of the high energy state. In the algorithm studied, the this state corresponds to the bottom of the sphere, with the low energy state being at the top. It can be assumed that the majority of this noise is occurring in the vertical direction only, as the qubit tends towards the $|0\rangle$ state. This is especially true near the end of the algorithm when the qubit states are close to the central axis. At this point the population of the $|1\rangle$ state is also very high. These considerations justify a further application of the previous extrapolation methods, where the procedures are applied only to the z coordinate of each point. Doing so produces the extrapolated trajectories of Figure \ref{fig:zonly}.

Figures \ref{fig:zonlyLinear} and \ref{fig:zonlyRich} can be compared to Figures \ref{fig:linear} and \ref{fig:richardson} of the previous sections to see how only extrapolating in the z direction changes the result. In both the z only cases the results produced by extrapolation are slightly smoother, specifically near the 20th and 23rd points of the trajectory. These are small improvements, but are significant due to their locations. The improvements are most prominent at points near the bottom of the Bloch sphere, where the noise is more strongly affecting the qubit, but also near the right and back sides of the Bloch sphere where decoherence can be important. The better results in these areas suggest that small amounts of decoherence were being over-corrected for in the three dimensional extrapolations. This indicates that decoherence may be more prominent along the x and y axis than the noise being studied. Considering these implications, it is reasonable to conclude that performing extrapolation in the z direction only offers a better, and more justified, noise mitigation in this case.

\section{Conclusions}
Current quantum computing hardware is noisy and error-prone, which often leads to poor measurements. Until the hardware is improved, error mitigation procedures will be required to obtain useful results from noisy devices. This study focused on mitigating noise in the IBM quantum experience by intentionally adding noise and performing extrapolation to zero noise. There are two conclusions to be drawn from this work: the ability to consistently add noise, and the effectiveness of extrapolation.

Noise was intentionally added  by using identity delays to extend the execution time of the computation. There were multiple ways to implement this, with three distinct methods considered. It was determined that a method where the delays are placed between each other operation resulted in a consistent increase in noise. This demonstrates that noise can be intentionally added to a quantum computation in a controlled manner.

With the intentionally noisy results two types of extrapolation were performed: linear and Richardson. While both methods did improve the results, it was suggested that only the Richardson procedure be considered for application in more complex cases. This is because the linear method required the exact solution to be known, while Richardson did not. In addition, it was determined that for the specific case studied here, performing extrapolation only in the vertical coordinate of the qubit was more justified and offered a better extrapolated result. 

The numerical results in Figure \ref{fig:richardson} show that the use of delay pulses for noise injection enabled extremely effective zero-noise extrapolation. To compare these results to the stretched gate method, see Figure 2(b) of \cite{kandala}. While it is clear that the delay pulse method achieved considerably more noise mitigation, one should note that the baseline trajectory (without injected noise) was considerably less noisy than in \cite{kandala}. 

As a final point, we mention that the techniques of this work are not a solution to quantum hardware issues, but rather a possible method to post-process quantum computation results in order to increase their reliability.

\section*{Acknowledgment}
We thank M. Amin, H. M{\"u}ller, T. Tiedje, and T. Zaborniak for encouragement and useful discussions during the execution of this work.

\end{document}